\documentclass[12pt,a4paper]{article}

\usepackage{lineno}  
\usepackage{listings}
\usepackage{graphicx}  

\usepackage{xspace}
\usepackage{color}
\usepackage{colortbl}
\usepackage{amsmath}
\usepackage{rotating}
\usepackage{listings}

\usepackage{ifthen} 

\usepackage[pdftex,
            plainpages=false,
            bookmarks  = true,
	    bookmarksnumbered = true,
	    pdfpagemode       = None,
	    pdfstartview      = FitH,
            pdfpagelayout     = SinglePage,
	    colorlinks        = true,
	    linkcolor         = blue,
	    pagecolor         = black,
	    urlcolor          = blue,
	    pdfborder         = {0 0 0}
	    ]{hyperref}%
\newboolean{pdflatex}
\setboolean{pdflatex}{true} 

\newboolean{uprightparticles}
\setboolean{uprightparticles}{false} 
\newboolean{articletitles}
\setboolean{articletitles}{true} 

\usepackage{amssymb}
\usepackage{amsfonts}
\usepackage{upgreek}

\usepackage{hyperref}    
\usepackage[all]{hypcap} 

\usepackage{lhcb-symbols-def}

\usepackage{mciteplus}

\begin{document}

\begin{titlepage}
\pagenumbering{roman}

\vspace*{-1.5cm}
\centerline{\large EUROPEAN ORGANIZATION FOR NUCLEAR RESEARCH (CERN)}
\vspace*{1.5cm}
\hspace*{-0.5cm}
\begin{tabular*}{\linewidth}{lc@{\extracolsep{\fill}}r}
\ifthenelse{\boolean{pdflatex}}
{\vspace*{-2.7cm}\mbox{\!\!\!\includegraphics[width=.14\textwidth]{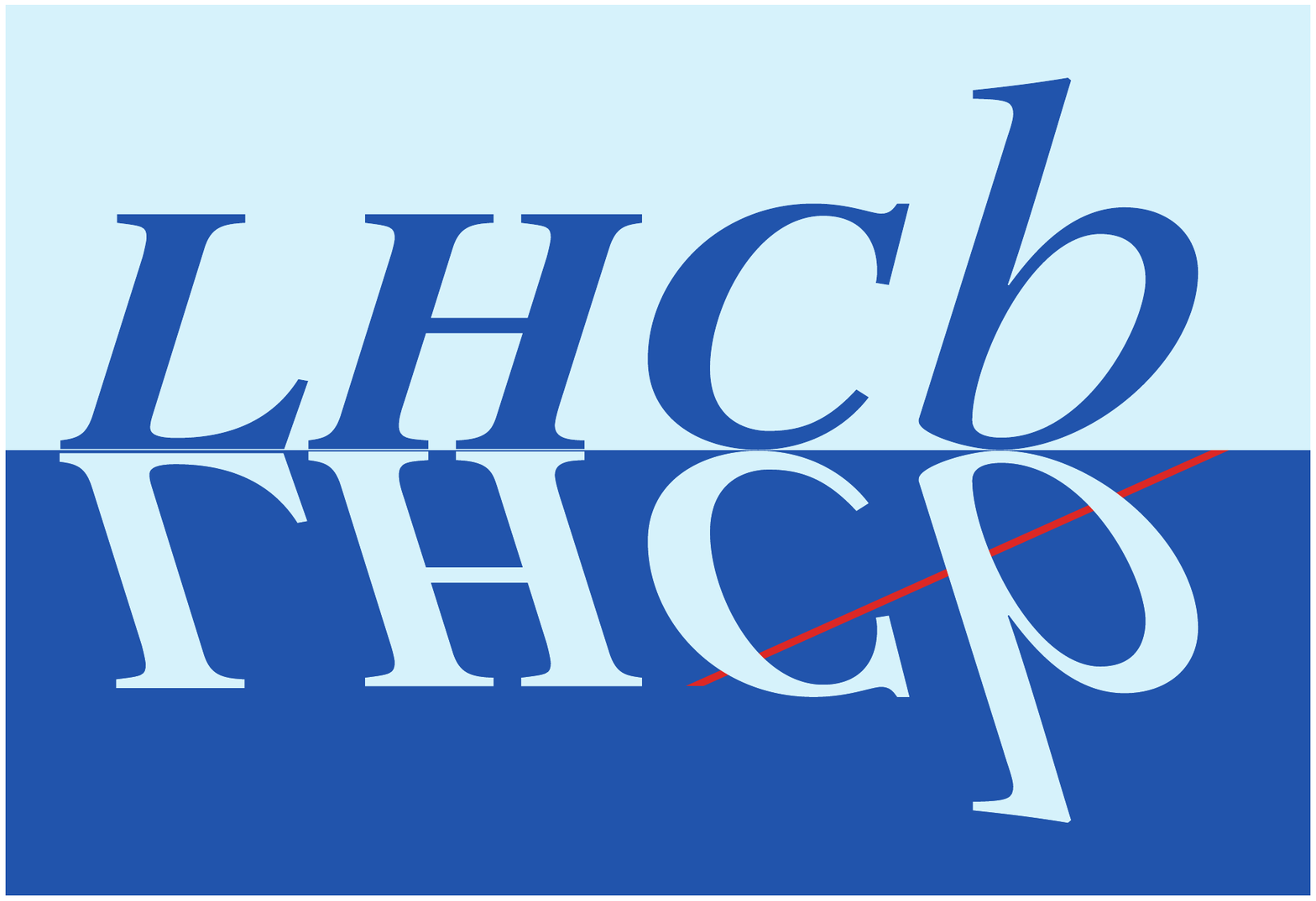}} & &}%
{\vspace*{-1.2cm}\mbox{\!\!\!\includegraphics[width=.12\textwidth]{lhcb-logo.eps}} & &}%
\\
 & & LHCb-PAPER-2011-035 \\  
 & & CERN-PH-EP-2011-226 \\  
 & & December 20, 2011; rev.\ January 25, 2012 \\ 
 & &        \\ 

\end{tabular*}

\vfill\vfill\vfill

{\bf\boldmath\LARGE
\begin{center}
Measurement of $b$-hadron masses \\\end{center}
}

\vfill

\begin{center}	
The LHCb Collaboration%
\footnote{Authors are listed on the following pages.}
\end{center}

\vfill

\vspace{\fill}

\begin{abstract}
\noindent
Measurements of $b$-hadron masses are performed with the exclusive
 decay modes $B^+\to J/\psi K^{+}$, \BdJKst\  , $B^0\to J/\psi K^{0}_{\rm S}$, $B_s^0\to J/\psi\phi$ and
$\Lambda^0_b\to J/\psi\Lambda$ using an integrated luminosity of
$35~\mathrm{pb}^{-1}$ collected in $pp$ collisions at a centre-of-mass energy of 7\tev
by the LHCb experiment. 
The momentum scale is calibrated with $J/\psi \to \mu^+\mu^-$ decays
and verified to be known to a relative precision of 2 $\times 10^{-4}$ using other
two-body decays. The results are more precise than previous
measurements, particularly in the case of the $B^0_s$ and $\Lambda^0_b$ masses.
\end{abstract}

\begin{center}
\textit{Published in Phys.\ Lett.\ B 708 (2012) 241--248}
\end{center}

\vspace{\fill}

\end{titlepage}

\newpage
\setcounter{page}{2}
\mbox{~}
\newpage

\begin{flushleft}
\centerline{\large\bf The LHCb Collaboration} 
\vspace{2ex}
\small 
R.~Aaij$^{23}$, 
C.~Abellan~Beteta$^{35,n}$, 
B.~Adeva$^{36}$, 
M.~Adinolfi$^{42}$, 
C.~Adrover$^{6}$, 
A.~Affolder$^{48}$, 
Z.~Ajaltouni$^{5}$, 
J.~Albrecht$^{37}$, 
F.~Alessio$^{37}$, 
M.~Alexander$^{47}$
% ...

{\footnotesize \it
% ...
$ ^{5}$Clermont Universit\'{e}, Universit\'{e} Blaise Pascal, CNRS/IN2P3, LPC, Clermont-Ferrand, France\\
$ ^{6}$CPPM, Aix-Marseille Universit\'{e}, CNRS/IN2P3, Marseille, France\\
% ...
$ ^{23}$Nikhef National Institute for Subatomic Physics, Amsterdam, The Netherlands\\
% ...
$ ^{35}$Universitat de Barcelona, Barcelona, Spain\\
$ ^{36}$Universidad de Santiago de Compostela, Santiago de Compostela, Spain\\
$ ^{37}$European Organization for Nuclear Research (CERN), Geneva, Switzerland\\
% ...
$ ^{42}$H.H. Wills Physics Laboratory, University of Bristol, Bristol, United Kingdom\\
% ...
$ ^{47}$School of Physics and Astronomy, University of Glasgow, Glasgow, United Kingdom\\
$ ^{48}$Oliver Lodge Laboratory, University of Liverpool, Liverpool, United Kingdom\\
% ...
$ ^{n}$LIFAELS, La Salle, Universitat Ramon Llull, Barcelona, Spain\\
% ...
}
\bigskip
\end{flushleft}

\cleardoublepage


\pagestyle{plain} 
\setcounter{page}{1}
\pagenumbering{arabic}


\section{Introduction} \label{sec:introduction}

Within the Standard Model of particle physics, mesons and baryons are colourless objects
composed of quarks and gluons.
These systems are bound through the strong interaction, described by quantum chromodynamics (QCD). 
A basic property of hadrons that can be compared to theoretical predictions is their masses.
The most recent theoretical predictions based on lattice QCD calculations can be found in Refs.~\cite{BsTheo,LambdabTheo}. 
The current experimental knowledge of the $b$-hadron masses as
summarized in Ref.~\cite{Nakamura:2010zzi} is dominated by results
from the CDF collaboration~\cite{CDFMasses}. 
In this Letter precision measurements of the masses of the $B^+$,
$B^0$, $B_s^0$ and $\Lambda^0_b$ are presented as well as the mass
splittings with respect to the $B^+$. 
The results are based on a data sample of proton-proton collisions at $\sqrt{s}=7\tev$ at the
Large Hadron Collider collected by the LHCb experiment, corresponding to an integrated luminosity of $\rm 35~pb^{-1}$.

The LHCb detector~\cite{LHCbDetector} is a forward spectrometer providing 
charged particle reconstruction in the pseudorapidity range 
$2<\eta<5$. The most important elements for the analysis presented
here are precision tracking and excellent particle
identification. The tracking system consists of a silicon strip vertex
detector (VELO) surrounding the $pp$ interaction region, a large area
silicon strip detector located upstream of a dipole magnet with a bending power of about 4~Tm,
and a combination of silicon strip detectors and straw drift-tubes placed downstream.
The combined tracking system has 
a momentum resolution $\delta p/p$ that varies from 0.4\% at
5\gevc to 0.6\% at 100\gevc. Pion, kaon and proton separation is provided by two ring imaging Cherenkov (RICH) 
detectors whilst muons are
identified by a muon system consisting of alternating layers of
iron and multi-wire proportional chambers. 

The data used for this analysis were collected in 2010. The trigger
system consists of two levels. The first stage is implemented in
hardware and uses information from the calorimeters and the muon
system. The second stage is implemented in software and runs on an
event filter farm. Dedicated trigger lines collect events containing 
$J/\psi$ mesons. For this analysis all events are used regardless of 
which trigger line fired. 

Simulation samples used are based on the {\sc{Pythia}}~6.4 generator~\cite{pythia} configured with the parameters detailed in Ref.~\cite{belyaev}. QED final state radiative
corrections are included using the {\sc{Photos}} package~\cite{photos}.
The {\sc{EvtGen}}~\cite{evtgen} and  {\sc{Geant4}}~\cite{geant} packages are used 
to generate hadron 
decays and simulate interactions in the detector, respectively. 

\begin{figure}[t]
\begin{center}
\includegraphics*[width=1\textwidth,viewport=32 0 565 122]{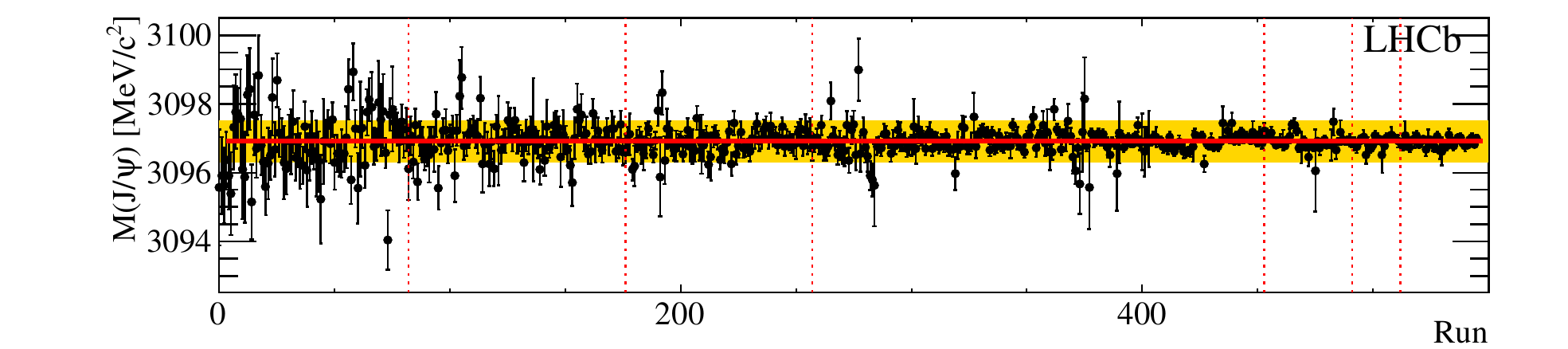}
\caption{\small Reconstructed $J/\psi\to\mu^+\mu^-$ fitted mass as a function of run number
after the momentum calibration procedure discussed in the text. 
The vertical dashed lines indicate the boundaries of the seven calibration periods.
A fit of a constant function (horizontal line) has a $\chi^2$ probability of $6\%$. 
The shaded area corresponds to the assigned uncertainty on the momentum scale of $0.02\%$.}
\label{jpsistability}
\end{center}
\end{figure}

\begin{figure}[t]
\begin{center}
\includegraphics*[width=0.6\textwidth,viewport=78 20 540 340]{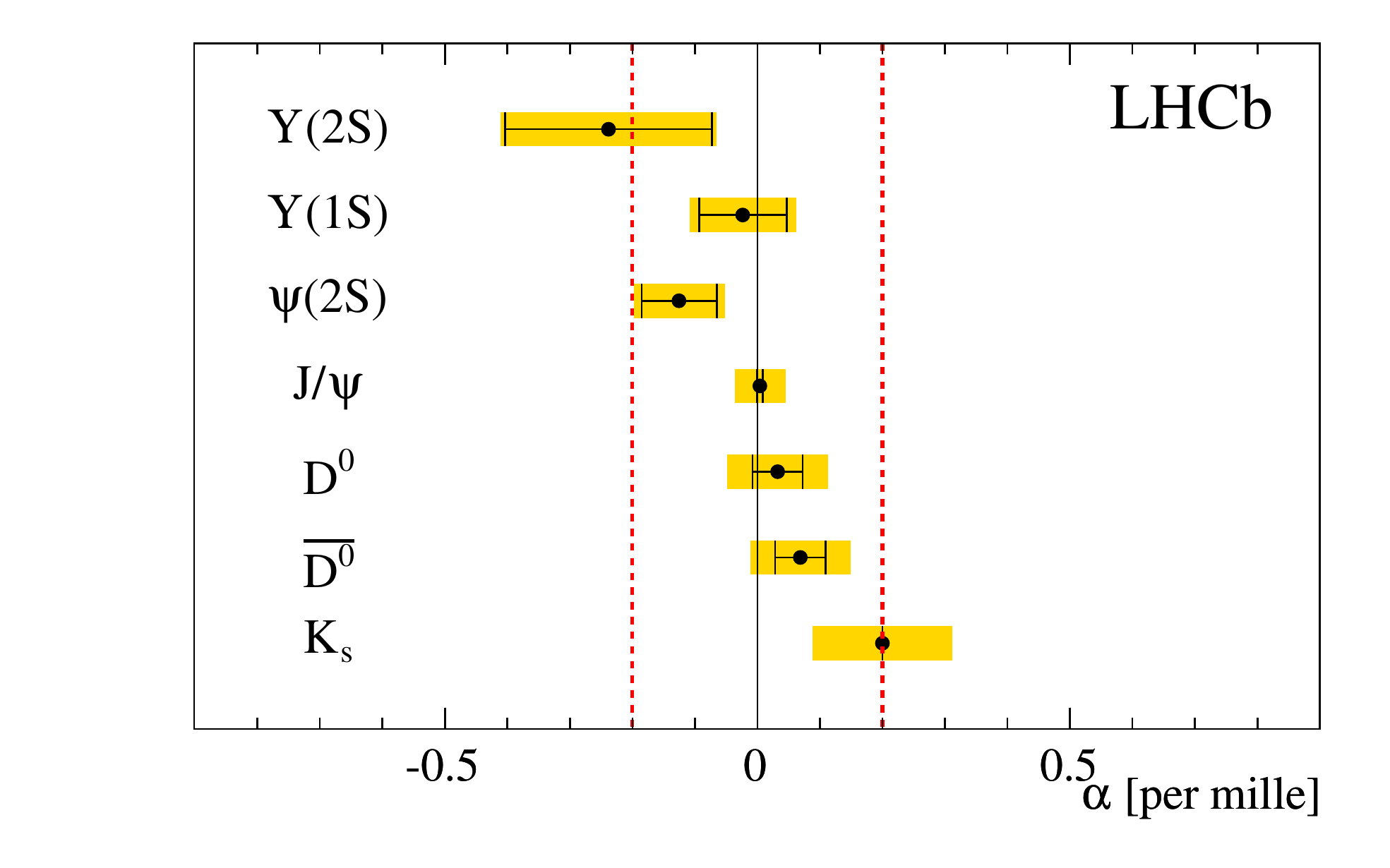}
\caption{\small Momentum scale bias $\alpha$, extracted from the reconstructed mass 
of various two-body decays
after the momentum calibration procedure described in the text. 
By construction one expects
$\alpha=0$ for the $J/\psi\to\mu^+\mu^-$ calibration mode. 
The black error bars represent the statistical
uncertainty whilst the (yellow) shaded areas include contributions to the 
systematic error from the fitting
procedure, the effect of QED radiative corrections and the
uncertainty quoted by the PDG~\cite{Nakamura:2010zzi}~on the mass of
the decaying meson. The (red) dashed lines correspond to the assigned uncertainty on the momentum scale of $0.02\%$. }
\label{alphax}
\end{center}
\end{figure}

\label{sec:scale}
The alignment of the tracking system, as well as the calibration of the momentum scale 
based on the $J/\psi \to \mu^+\mu^-$ mass peak,
were carried out in seven time periods corresponding to different
running conditions. The procedure takes into account the effects of QED radiative
corrections which are important in the $J/\psi \to \mu^+\mu^-$ decay.
Figure~\ref{jpsistability}  shows that the reconstructed $J/\psi$ mass after alignment and calibration 
is stable in time to better than $0.02\%$ throughout the data-taking
period. The validity of the momentum calibration has been checked using samples of
$K^0_{\rm S}\to \pi^+\pi^-$, $D^0 \to K^-\pi^+$, $\bar{D}^0 \to K^+\pi^-$, 
$\psi(2S)\to \mu^+\mu^-$, $\Upsilon(1S)\to \mu^+\mu^-$ and $\Upsilon(2S)\to \mu^+\mu^-$ decays. 
In each case the mass distribution is modelled taking
into account the effect of radiative corrections, resolution and background, 
and the mean mass value extracted. To allow comparison between the
decay modes, the deviation of the measured mass 
from the expected value~\cite{Nakamura:2010zzi} is converted
into an estimate of the momentum scale bias, referred to as
$\alpha$. This is defined such that the measured mass is equal to the expected 
value if all particle momenta are multiplied by $1-\alpha$.
Figure~\ref{alphax} shows the resulting values of $\alpha$. The 
deviation for the considered modes is $\pm 0.02\%$, which is taken as the 
systematic uncertainty on the momentum scale. 

\section{Event selection} \label{sec:selection}

A common strategy, aiming
at high signal purity, is adopted for the
reconstruction and selection of 
$B^+\to J/\psi K^{+}$, \BdJKst, $B^0\to J/\psi K^0_{\rm S}$, $B_s^0\to J/\psi\phi$ and
$\Lambda^0_b\to J/\psi\Lambda$ candidates (the inclusion of
charge-conjugated modes is implied throughout). In general, only tracks traversing the whole spectrometer are used;  
however, since $K^0_{\rm S}$ and $\Lambda$ particles may decay outside of the 
VELO, pairs of tracks without VELO hits are also used to build
$K^0_{\rm S}$ and $\Lambda$ candidates.  The $\chi^2$ per number of
degrees of freedom ($\chi^2/{\rm ndf}$) of the track fit is required
to be smaller than four. The Kullback-Leibler (KL) distance~\cite{kl,*kl2,*kl3}
is used to identify pairs of reconstructed tracks that are very likely to arise
from hits created by the same charged particle: 
if two reconstructed tracks have a symmetrized KL divergence less than 5000,
only that with the higher fit quality is considered.

$J/\psi \to \mu^+\mu^-$ candidates are formed from pairs of oppositely-charged  muons with a transverse momentum ($p_{\rm T}$) larger
than $0.5\gevc$, originating from a common vertex with $\chi^2/{\rm ndf}< 11$, 
and satisfying $|M_{\mu\mu}-M_{J/\psi}| < 3\sigma$
where $M_{\mu\mu}$ is the reconstructed dimuon mass, $M_{J/\psi}$ is the known $J/\psi$ mass world average value~\cite{Nakamura:2010zzi},
and $\sigma$ is the estimated event-by-event uncertainty on $M_{\mu\mu}$.
The selected $J/\psi$ candidates are then combined with one of $K^+$, $K^{*0} \to K^+\pi^-$, $\phi\to K^+K^-$, $K^0_{\rm S} \to
\pi^+\pi^-$ or $\Lambda \to p \pi^-$ to create $b$-hadron candidates.
Mass windows of $\pm 70\mevcc$, $\pm 12\mevcc$, 
$\pm 12\mevcc$ ($\pm 21\mevcc$) and $\pm 6\mevcc$ ($\pm 6\mevcc$) 
around the world averages~\cite{Nakamura:2010zzi}
are used to select the $K^{*0}$, $\phi$, $K_{\rm S}^0$ and $\Lambda$ candidates
formed from tracks with (without) VELO hits, respectively. Kaons are selected by cutting on the
difference between the log-likelihoods of the kaon and pion
hypotheses provided by the RICH
detectors ($\Delta \ln{\cal L}_{K-\pi} > 0$). To eliminate background
from $B_s^0 \to J/\psi \phi$ in the $B^0 \to
J/\psi K^{*0}$ channel, the pion from the $K^{*0}$
candidate is required to be inconsistent with the kaon hypothesis
($\Delta  \ln{\cal L}_{K-\pi} < 0$).
To further improve the signal purity, a requirement of $\pT > 1\gevc$ 
is applied on the particle associated with the $J/\psi$ candidate. For final states including a $V^0$ ($K_{\rm S}^0$ or $\Lambda$), 
an additional requirement of $L / \sigma_L > 5$ is made, where $L$ is
the distance between the $b$-hadron and the $V^0$ decay vertex,
and $\sigma_L$ is the uncertainty on this quantity.  

Each $b$-hadron candidate is associated with the reconstructed $pp$
primary interaction vertex with respect to which it has the smallest
impact parameter significance, and this significance is required
to be less than five. As there is a large combinatorial background due to particles
originating directly from the $pp$ primary vertex, only $b$-hadron candidates
with a reconstructed decay time greater than 0.3~ps are considered for subsequent
analysis. 
A decay chain fit~\cite{Hulsbergen} is performed for each candidate, 
which constrains the reconstructed 
$J/\psi$ mass and, if applicable, the reconstructed $K_{\rm S}^{0}$ or $\Lambda$ mass 
to their nominal values~\cite{Nakamura:2010zzi}. The $\chi^2/{\rm ndf}$ of the
fit is required to be smaller than five. The mass of the $b$-hadron
candidate is obtained from this fit and its estimated uncertainty is required to be 
smaller than 20\mevcc. 

\section{Results} \label{sec:fitting}

\begin{figure}[t]
\vspace{-0.5ex}
\begin{center}
\includegraphics[scale=0.35]{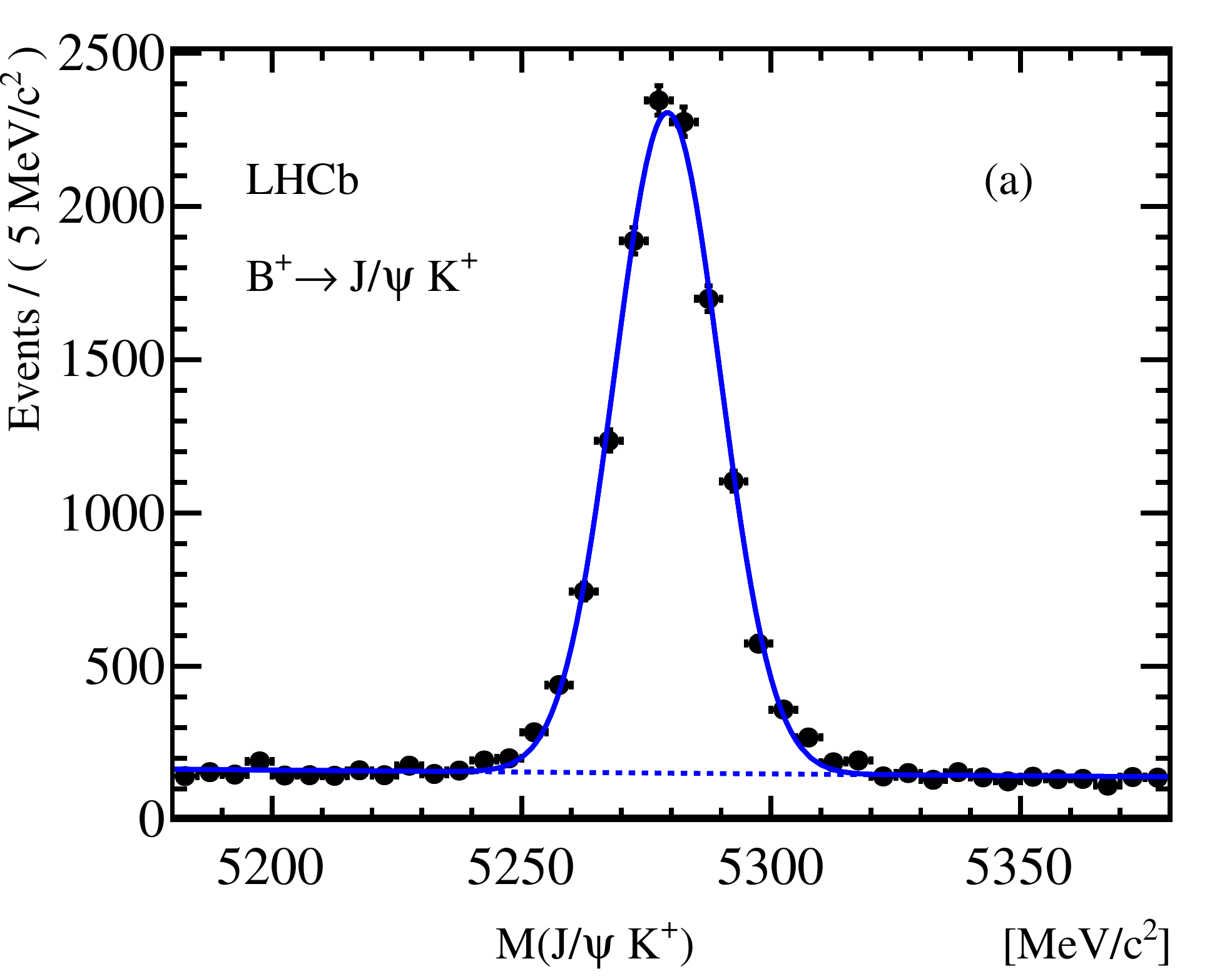} \includegraphics[scale=0.35]{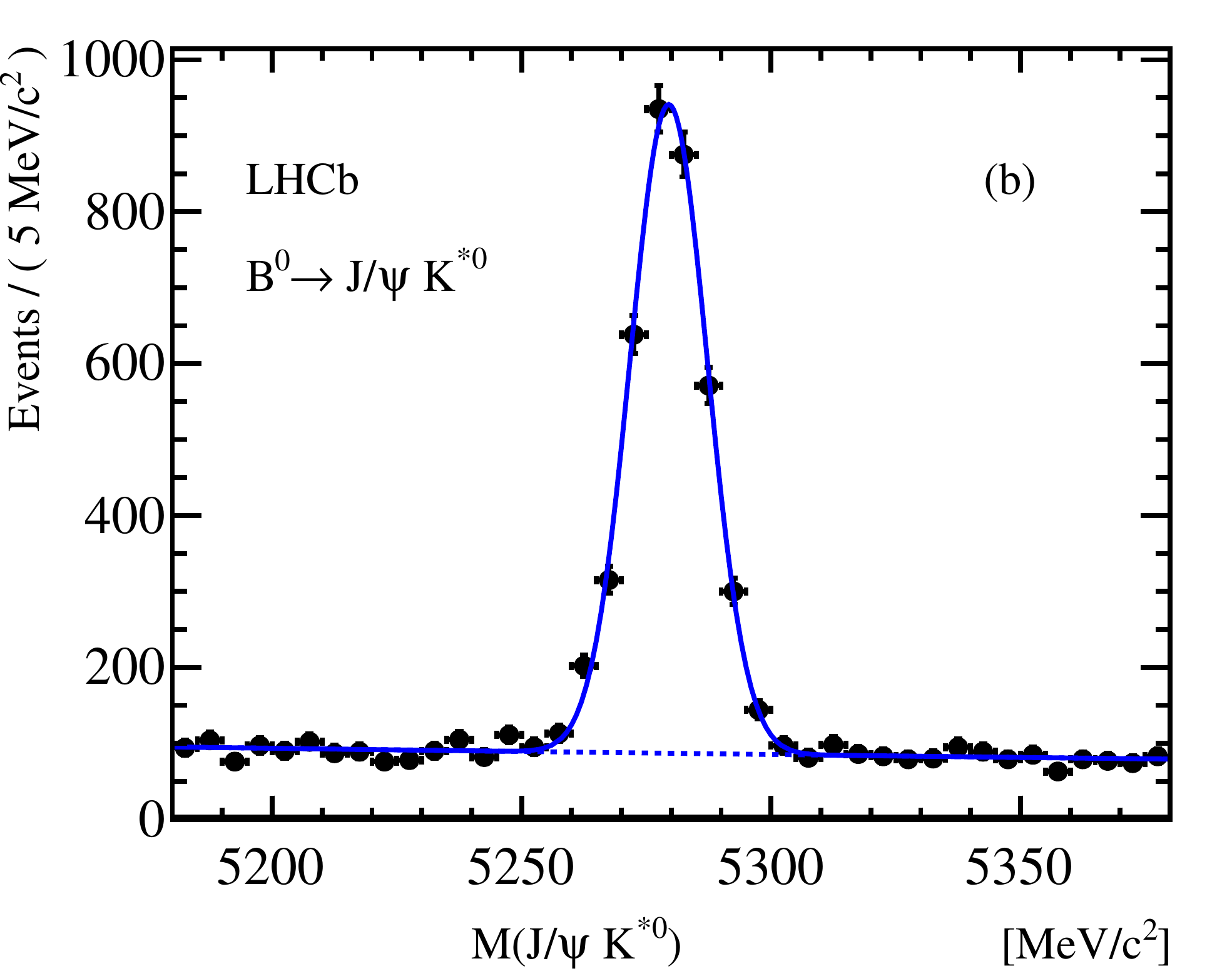} \\
\includegraphics[scale=0.35]{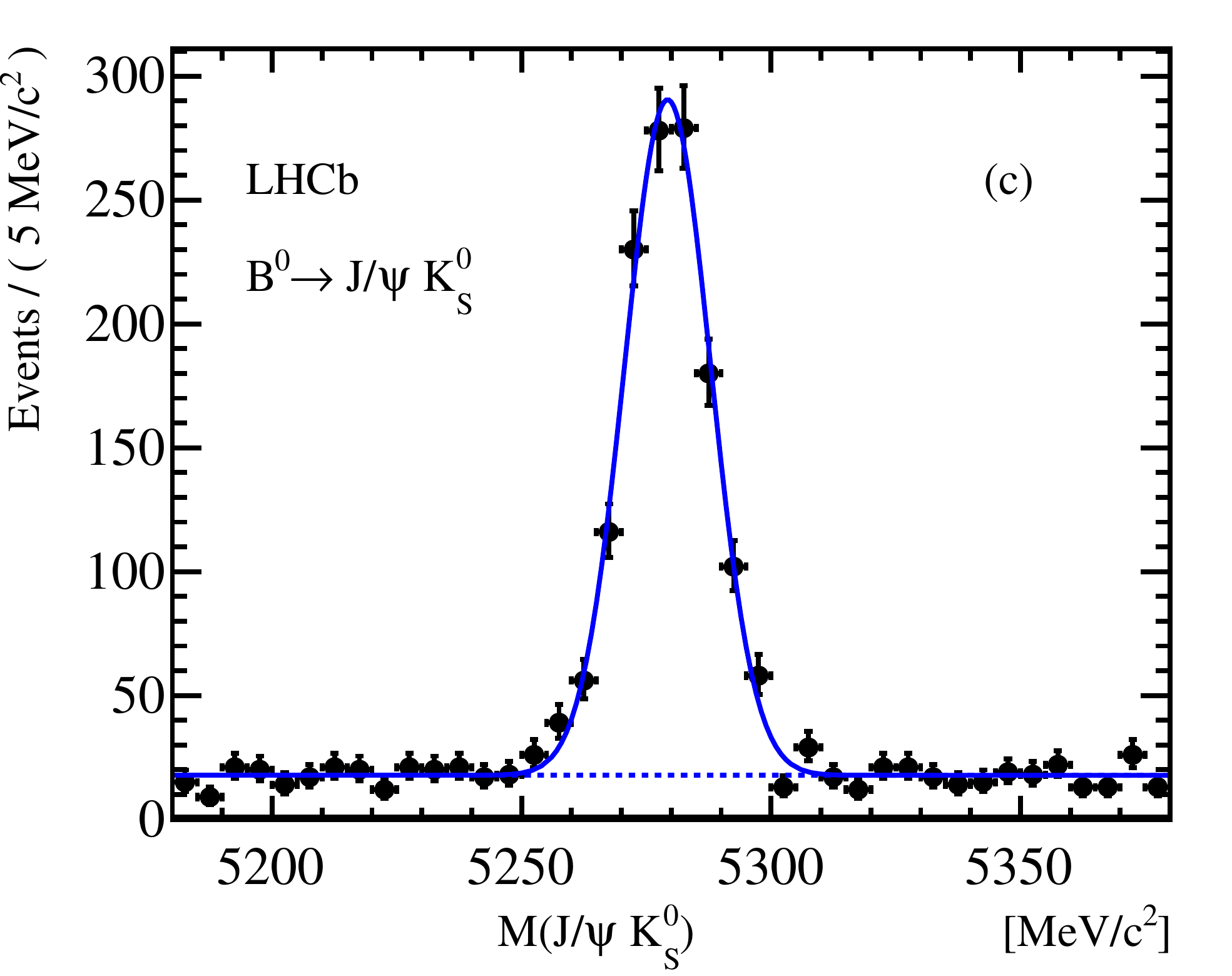} \includegraphics[scale=0.35]{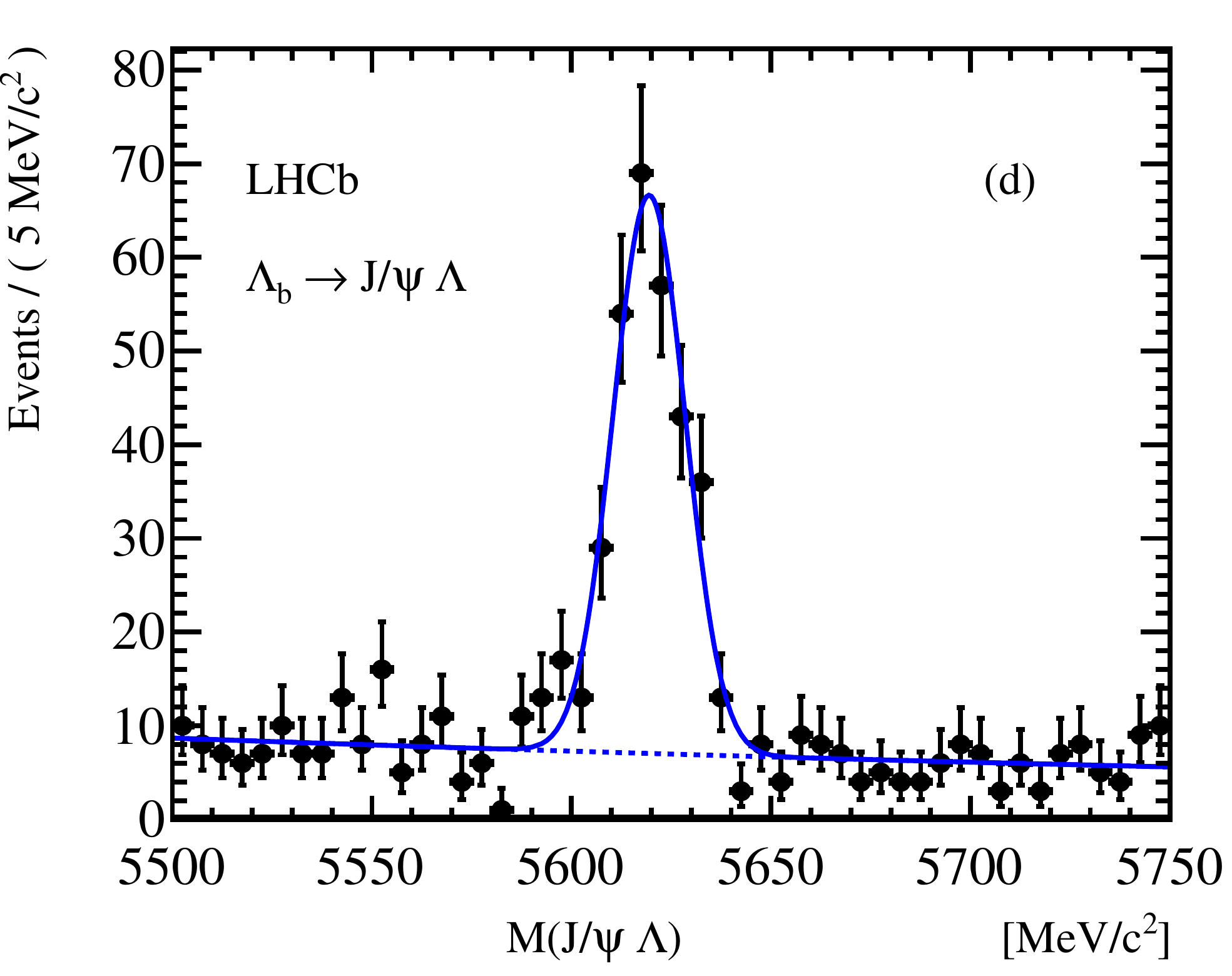} \\
\includegraphics[scale=0.35]{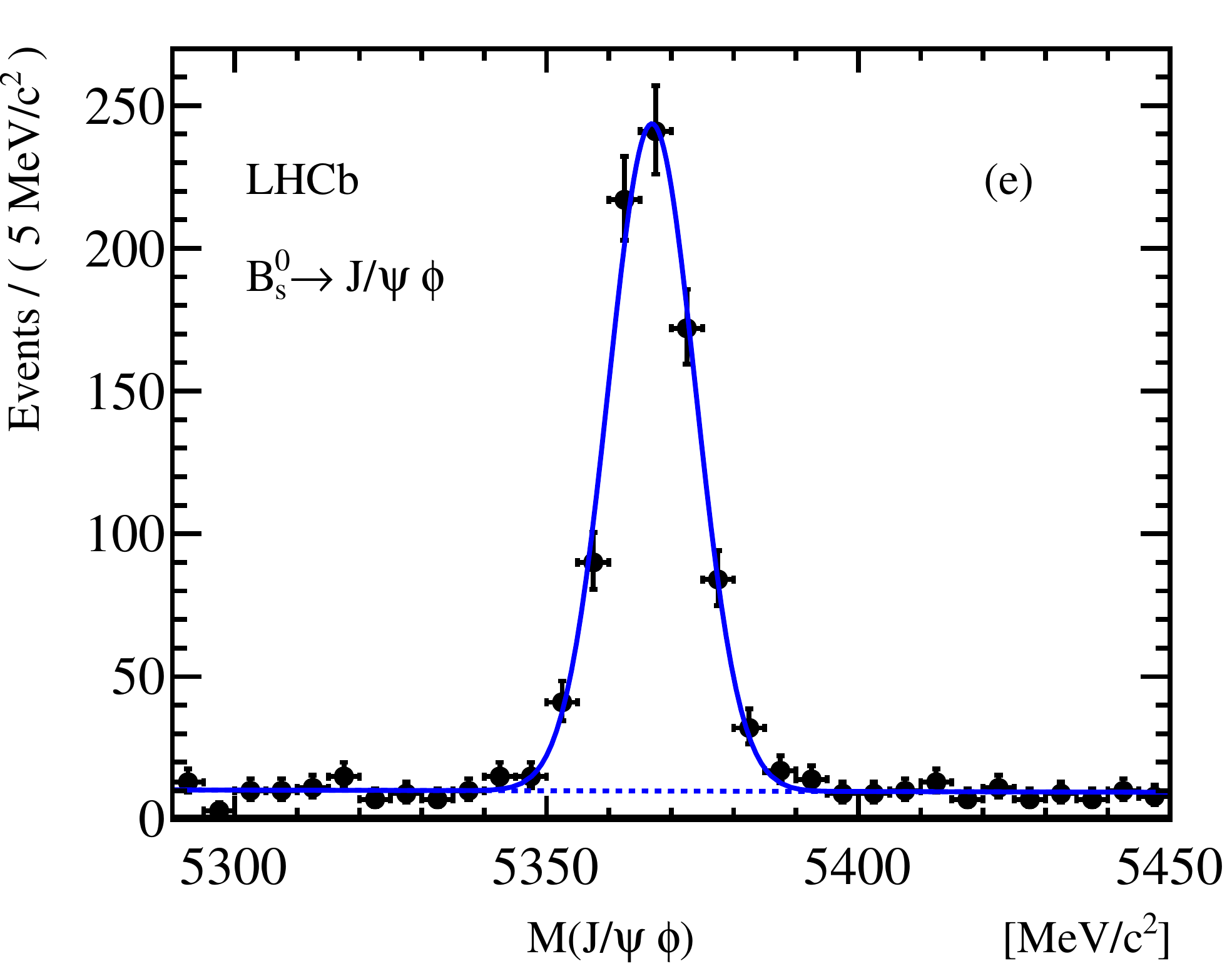}
\end{center}
\vspace{-3.5ex}
\caption{\small Invariant mass distributions for  
(a) $B^+\to J/\psi K^+$, (b) $B^0\to J/\psi K^{*0}$,
(c) $B^0\to J/\psi K^0_{\rm S}$, (d) $ \Lambda^0_b\to J/\psi\Lambda$, and
(e) $B_s^0\to J/\psi\phi$ candidates. In each case the result of the fit
described in the text is superimposed (solid line) together with the
background component (dotted line).}
\label{fig:Masses}
\end{figure}

The $b$-hadron masses are determined by performing unbinned maximum
likelihood fits to the invariant mass distributions, in which the 
signal and background components are described by a Gaussian and an exponential function, respectively. 
Alternative models for both the signal and
background components are considered as part of the systematic studies.
Figure~\ref{fig:Masses} shows the invariant mass
distributions and fits for the five modes considered in this
study. The signal yields, mass values and resolutions resulting from the
fits are given in Table~\ref{tab:yields}.  

\begin{table}[t]
\caption{\small Signal yields, mass values and mass resolutions
  obtained from the fits shown in Fig.~\ref{fig:Masses} together with
  the values corrected for the effect of QED radiative corrections as
  described in the text. The quoted uncertainties are statistical.}
\begin{center}
\begin{tabular}{l|c|c|c|c}
 &  &  Fitted mass  &  Corrected mass  & Resolution \\ 
 \raisebox{1.5ex}[-1.5ex]{Decay mode} &
 \raisebox{1.5ex}[-1.5ex]{Yield} &[$\!\mevcc$] & [$\!\mevcc$]  &  [$\!\mevcc$] \\ \hline
$B^+\to J/\psi K^+$            & $11151     \pm 115$ &   $5279.24 \pm 0.11$   & $5279.38 \pm 0.11$ & $10.5   \pm 0.1$ \\
$B^0\to J/\psi K^{*0}$         & $\,~3308   \pm \,~65 $ & $5279.47 \pm 0.17$ & $5279.58 \pm 0.17$ & $\,~7.7 \pm 0.2$ \\
$B^0\to J/\psi K^0_{\rm S}$    & $\,~1184   \pm \,~38 $ &  $5279.58 \pm 0.29$ & $5279.58 \pm 0.29$ & $\,~8.6 \pm 0.3$ \\
$B_s^0\to J/\psi\phi$          & $\,~\,~816 \pm \,~30 $ &$5366.90 \pm 0.28$ & $5366.90 \pm 0.28$ & $\,~7.0 \pm 0.3$ \\
$\Lambda^0_b\to J/\psi\Lambda$ & $\,~\,~279 \pm \,~19 $ &  $5619.19
\pm 0.70$  & $5619.19 \pm 0.70$ & $\,~9.0 \pm 0.6$ 
\end{tabular}
\end{center}
\label{tab:yields}
\end{table}

The presence of biases due to neglecting QED radiative corrections in
the mass fits is studied using a simulation based on {\sc
  Photos}~\cite{photos}.  The fitted masses quoted in Table~\ref{tab:yields} for the $B^+\to J/\psi K^+$
and $B^0\to J/\psi K^{*0}$ are found to be underestimated by $0.14 \pm 0.01\mevcc$
and $0.11 \pm 0.01\mevcc$, respectively, when radiative corrections are ignored; they are therefore corrected 
for these biases, and the uncertainty is propagated as a systematic effect.
The bias for the $B^0_s \to J/\psi \phi$ mode is negligible due to the restricted
phase space for the kaons from the $\phi$ decay. There is no bias for
the $B^0 \to J/\psi K^0_{\rm S}$ and $\Lambda^0_b \to J/\psi \Lambda$
modes since the $J/\psi$, $K^0_{\rm S}$ and $\Lambda$ masses are constrained in the vertex fits. 

\section{Systematic studies and checks} \label{sec:systematics}

To evaluate the systematic error, the complete analysis is repeated (including the track fit and the
momentum scale calibration when needed), varying within their
uncertainties the parameters to which the mass determination is
sensitive. The observed changes in the central values of the 
fitted masses relative to the nominal results are then assigned as systematic
uncertainties. 

The dominant source of uncertainty is  the
limited knowledge of the momentum scale. The mass fits are repeated with the momentum
scale varied by $\pm 0.02 \%$. After the calibration procedure a $\pm
0.07\%$ variation of the momentum scale remains as a function of the
particle pseudorapidity $\eta$. To first order the effect of this
averages out in the mass determination. The residual impact of this variation is evaluated by parameterizing the
momentum scale as a function of $\eta$ and repeating the analysis. 
The amount of material traversed in the tracking system by a particle is 
known to 10\% accuracy~\cite{Aaij:2010nx}; the magnitude of the energy 
loss correction in the reconstruction is therefore varied by 10\%. 
To ensure the detector alignment is well understood a
further test is carried out: the horizontal and vertical
slopes of the tracks close to the
interaction region, which are determined by measurements in the VELO,
are changed by 1$\times10^{-3}$, corresponding to 
the precision with which the length scale along the beam axis is
known~\cite{dmspaper}. Other uncertainties arise from the fit
modelling: a double Gaussian function (with common mean) for the signal resolution 
and/or a flat background component are used instead
of the nominal Gaussian and exponential functions. 
The effect of possible reflections due to particle mis-identification
is small and can be neglected.
Finally, a systematic uncertainty related to the evaluation of the
effect of the radiative corrections is assigned.  
Tables~\ref{tab:Syst} and \ref{tab:SystDiff} summarize the systematic uncertainties
assigned on the measured masses and mass differences. 

\begin{table}[t]
\caption{\small Systematic uncertainties (in \MeVcc) on the mass measurements.}
\begin{center}
\begin{tabular}{l|r|r|r|r|r}
Source of uncertainty 
& \multicolumn{1}{@{}c@{}|}{$B^+ \to$}
& \multicolumn{1}{@{}c@{}|}{$B^0 \to$}
& \multicolumn{1}{@{}c@{}|}{$B^0 \to$}
& \multicolumn{1}{@{}c@{}|}{$B_s^0 \to$}
& \multicolumn{1}{@{}c@{}}{$\Lambda^0_b \to$}\\
& \multicolumn{1}{@{\,}c@{\,}|}{$J/\psi K^+$}
& \multicolumn{1}{@{\,}c@{\,}|}{$J/\psi K^{*0}$}
& \multicolumn{1}{@{\,}c@{\,}|}{$J/\psi K^0_{\rm S}$}
& \multicolumn{1}{@{\,}c@{\,}|}{$J/\psi \phi$}
& \multicolumn{1}{@{\,}c@{\,}}{$J/\psi \Lambda$}\\
\hline 
{Mass fitting:}&&&&& \\ 
~ -- Background model                    & 0.04 &    0.03 & $<$0.01 & 0.01 & $<$0.01 \\ 
~ -- Resolution model                    & 0.01 &    0.02 &    0.06 & 0.02 &    0.07 \\  
~ -- Radiative corrections               & 0.01 &    0.01 &      -- &   -- &     --  \\  
{Momentum calibration:}&&&&& \\
~ -- Average momentum scale              & 0.30 &    0.27 &    0.30 & 0.22 &    0.27 \\ 
~ -- $\eta$ dependence of momentum scale & 0.04 & $<$0.01 &    0.09 & 0.03 &    0.02 \\ 
{Detector description:}&&&&& \\
~ -- Energy loss  correction             & 0.10 & $<$0.01 &    0.05 & 0.03 &    0.09 \\ 
{Detector alignment:}&&&&& \\
~ -- Vertex detector (track slopes)      & 0.05 &    0.04 &    0.04 & 0.03 &    0.04 \\ 
\hline
Quadratic sum                            & 0.33 &    0.27 &    0.33 & 0.23 &    0.30 \\
\end{tabular}
\end{center}
\label{tab:Syst}
\caption{\small Systematic uncertainties (in \MeVcc) on the differences of mass measurements, 
expressed with respect to the $B^+ \to J/\psi K^+$ mass
(\textit{e.g.} the last column gives the systematic uncertainties on
$M(\Lambda_b^0 \to J/\psi\Lambda)-M(B^+ \to J/\psi K^+)$).}
\begin{center}
\begin{tabular}{l l|r|r|r|r}
Source of uncertainty 
& \multicolumn{1}{@{}c@{}|}{~~~~~~~~~}
& \multicolumn{1}{@{}c@{}|}{$B^0 \to$}
& \multicolumn{1}{@{}c@{}|}{$B^0 \to$}
& \multicolumn{1}{@{}c@{}|}{$B_s^0 \to$}
& \multicolumn{1}{@{}c@{}}{$\Lambda^0_b \to$}\\
& \multicolumn{1}{@{\,}c@{\,}|}{~~~}
& \multicolumn{1}{@{\,}c@{\,}|}{$J/\psi K^{*0}$}
& \multicolumn{1}{@{\,}c@{\,}|}{$J/\psi K^0_{\rm S}$}
& \multicolumn{1}{@{\,}c@{\,}|}{$J/\psi \phi$}
& \multicolumn{1}{@{\,}c@{\,}}{$J/\psi \Lambda$}\\
\hline 
{Mass fitting:}&&&&& \\ 
~ -- Background model                    &       &    0.05 &    0.04 & 0.04 & 0.04 \\ 
~ -- Resolution model                    &       &    0.02 &    0.06 & 0.02 & 0.07 \\  
~ -- Radiative corrections               &       & $<$0.01 &    0.01 & 0.01 & 0.01 \\  
{Momentum calibration:}&&&&& \\
~ -- Average momentum scale              &       &    0.03 & $<$0.01 & 0.08 & 0.03 \\ 
~ -- $\eta$ dependence of momentum scale &       &    0.04 &    0.05 & 0.01 & 0.02 \\
{Detector description:}&&&&& \\
~ -- Energy loss  correction             &       &    0.10 &    0.05 & 0.07 & 0.01 \\ 
{Detector alignment:}&&&&& \\
~ -- Vertex detector (track slopes)      &       &    0.01 &    0.01 & 0.02 & 0.01 \\ 
\hline
Quadratic sum                         & ~~~~~~~~ &    0.12 &    0.10 & 0.12 & 0.09 \\
\end{tabular}
\end{center}
\label{tab:SystDiff}
\end{table}

The stability of the measured $b$-hadron masses is studied by dividing
the data samples according to the polarity of the spectrometer magnet,
final state flavour (for modes where the final state is flavour specific),
as well as whether the $K^0_{\rm S}$ and $\Lambda$ daughter particles
have VELO hits. As a cross-check the analysis is repeated ignoring the
hits from the tracking station before the magnet. This leads to an
average shift in measured masses compatible with statistical
fluctuations. In addition, for the $B^+$ and $B^0$ modes where 
the event samples are sizable, the measurements are repeated in bins
of the $b$-hadron kinematic variables. 
None of these checks reveals a systematic bias. 

\section{Conclusions} \label{conclusions}

The $b$-hadron masses are measured using data collected in 2010 
at a centre-of-mass energy of $\sqrt{s} = 7$\tev.
The results are
\begin{center}
\begin{tabular}{l@{$\,\to\,$}l@{$~=~$}l@{$\,\,\pm\,$}l@{\,(stat) $\pm\,$}l@{\,(syst)\,\mevcc}l}
$M(B^+$       & $J/\psi K^{+})$         & 5279.38 & 0.11 & 0.33 & , \\
$M(B^0$       & $J/\psi K^{(*)0})$        & 5279.58 & 0.15 & 0.28 & , \\
$M(B_s^0$     & $J/\psi\phi)$           & 5366.90 & 0.28 & 0.23 & , \\
$M(\Lambda^0_b$ & $J/\psi\Lambda)$        & 5619.19 & 0.70 & 0.30 & , \\
\end{tabular}
\end{center}
where the $B^0$ result is obtained as a weighted average of 
$M(B^0 \to J/\psi K^{*0}) = 5279.58 \pm 0.17 \pm 0.27\mevcc$ and 
$M(B^0 \to J/\psi K^{0}_{\rm S}) = 5279.58 \pm 0.29 \pm 0.33\mevcc$ assuming all 
systematic uncertainties to be correlated, except those related to the
 mass model. The dominant systematic uncertainty is related to the knowledge of
the average momentum scale of the tracking system. It largely
cancels in the mass differences. We obtain
\begin{center}
\begin{tabular}{l@{$\,-M(B^+ \to J/\psi K^{+})=\,\,$}r@{$\,\,\pm\,$}l@{\,(stat) $\pm\,$}l@{\,(syst)\,\mevcc}c}
$M(B^0 \to J/\psi K^{(*)0})$      &   0.20 & 0.17 & 0.11 & , \\ 
$M(B^0_s \to J/\psi\phi)$         &  87.52 & 0.30 & 0.12 & , \\ 
$M(\Lambda^0_b \to J/\psi\Lambda )$ & 339.81 & 0.71 & 0.09 & , \\ 
\end{tabular}
\end{center}
where the $B^0$ result is a combination of 
$M(B^0 \to J/\psi K^{*0})-M(B^+ \to J/\psi K^+) = 0.20 \pm 0.20 \pm 0.12\mevcc$ and 
$M(B^0 \to J/\psi K^0_{\rm S})-M(B^+ \to J/\psi K^+) = 0.20 \pm 0.31 \pm 0.10\mevcc$ under the same 
hypothesis as above. 

As shown in Table~\ref{tab:summary}, our measurements
are in agreement with previous measurements~\cite{Nakamura:2010zzi,CDFMasses}.
Besides the difference between the $B^+$ and $B^0$ masses they are the
most accurate to date, with significantly improved precision
over previous measurements in the case of the $B^0_s$ and $\Lambda^0_b$ masses.

\clearpage

\begin{table}[t]
\caption{\small LHCb measurements, compared to both the best previous
measurements and the results of a global fit to available $b$-hadron mass data~\cite{Nakamura:2010zzi}.
The quoted errors include statistical and systematic uncertainties. All values are in\,\mevcc.}
\begin{center}
\begin{tabular}{c|r|c|c}
 & \multicolumn{1}{c|}{LHCb}  & Best previous &  \\ 
\raisebox{1.5ex}[-1.5ex]{Quantity} & \multicolumn{1}{c|}{measurement} & measurement & \raisebox{1.5ex}[-1.5ex]{PDG fit} \\
\hline
$M(B^+)$         & $5279.38 \pm 0.35$  & $5279.10 \pm 0.55$ ~\cite{CDFMasses}  & $5279.17 \pm 0.29$ \\
$M(B^0)$         & $5279.58 \pm 0.32$  & $5279.63 \pm 0.62$ ~\cite{CDFMasses}  & $5279.50 \pm 0.30$ \\
$M(B_s^{0})$     & $5366.90 \pm 0.36$  & $5366.01 \pm 0.80$ ~\cite{CDFMasses} & $5366.3~\, \pm 0.6~\,$ \\
$M(\Lambda^0_b)$ & $5619.19 \pm 0.76$  & $5619.7~\, \pm 1.7~\,$ ~\cite{CDFMasses} & ~~~~-- \\
$M(B^0) -M(B^+)$  &  $~~~0.20\pm 0.20 $ & $~~~~~0.33 \pm 0.06$ \cite{babar} & $~~~~0.33 \pm 0.06$ \\   
$M(B^0_s) -M(B^+) $ &   $87.52 \pm 0.32 $ & --   & --\\  
$M(\Lambda^0_b) -M(B^+) $ & $339.81 \pm 0.72 $ & -- & --   \\
\end{tabular}
\end{center}
\label{tab:summary}
\end{table}

\appendix \section*{Acknowledgements}

\noindent We would like to thank our colleague Adl\`{e}ne Hicheur who made, as a
member of our collaboration, significant contributions to the tracking
alignment algorithms and provided the first realistic version
of the magnetic field map. He is currently unable to continue his work,
and we hope that this situation will be resolved soon. We express our gratitude to our colleagues in the CERN
accelerator departments for the excellent performance of the LHC. We thank the
technical and administrative staff at CERN and at the LHCb institutes,
and acknowledge support from the National Agencies: CAPES, CNPq,
FAPERJ and FINEP (Brazil); CERN; NSFC (China); CNRS/IN2P3 (France);
BMBF, DFG, HGF and MPG (Germany); SFI (Ireland); INFN (Italy); FOM and
NWO (The Netherlands); SCSR (Poland); ANCS (Romania); MinES of Russia and
Rosatom (Russia); MICINN, XuntaGal and GENCAT (Spain); SNSF and SER
(Switzerland); NAS Ukraine (Ukraine); STFC (United Kingdom); NSF
(USA). We also acknowledge the support received from the ERC under FP7
and the Region Auvergne.

\bibliographystyle{LHCb}
\bibliography{refs}

\end{document}